\documentstyle[12pt,epsf]{article}

\begin{document}

\newcommand{\balpha}{{\mbox{\boldmath$\alpha$}}}
\newcommand{\bmu}{{\mbox{\boldmath$\mu$}}}
\newcommand{\bnu}{{\mbox{\boldmath$\nu$}}}
\newcommand{\be}{\begin{eqnarray}}
\newcommand{\ee}{\end{eqnarray}}
\newcommand{\la}{\langle}
\newcommand{\ra}{\rangle}
\newcommand{\bfx}{{\bf x}}
\newcommand{\bfy}{{\bf y}}
\newcommand{\az}{\alpha Z}
\newcommand{\bfz}{{\bf z}}
\newcommand{\bfn}{{\bf n}}

\begin{center}

{\Large \bf 
QED Effects in  Heavy Few-Electron Ions}
\end{center}
\begin{center}
{\large V.M. Shabaev, O.V. Andreev, A.N. Artemyev, S.S. Baturin,
A.A. Elizarov, Y.S. Kozhedub, N.S. Oreshkina,
 I.I. Tupitsyn, V.A. Yerokhin, O.M. Zherebtsov}
\end{center}
$$
\,
$$
\begin{center}
{\it Department of Physics, St.Petersburg State University},\\
 {\it Oulianovskaya 1, Petrodvorets, St.Petersburg 198504, Russia}
\vspace{0.5cm}
\end{center}
PACS number(s): 12.20.-m, 12.20.Ds, 31.30.Jv, 31.30.Gs

\begin{abstract}

Accurate calculations of the binding energies,
the hyperfine splitting, the bound-electron g-factor,
and the parity nonconservation effects in heavy few-electron
ions are considered. The calculations include the relativistic,
quantum electrodynamic (QED), electron-correlation, and nuclear
effects. The theoretical results are compared with available
experimental data. A special attention is focused on tests
of QED in a strong Coulomb field.

\end{abstract}

\section{Introduction}

Accurate calculations of heavy few-electron
ions must be performed in the framework of the rigorous
quantum electrodynamic (QED) formalism.
The basic methods of quantum electrodynamics were formulated
to the beginning of 1930's,
almost immediately after creation of quantum
mechanics. This theory provided description of such low-order 
processes as emission and absorption of photons and creation
and annihilation of electron-positron pairs. However, application
of this theory to some higher-order effects gave infinite results.
This problem remained unsolved until the late 1940's when Lamb 
and Retherford discovered the $2s-2p_{1/2}$ splitting (Lamb shift)
in hydrogen.
This discovery stimulated theorists to complete the creation
of QED since it was believed that this splitting is of quantum electrodynamic
origin. First evaluation of the Lamb shift was performed by Bethe who
used Kramer's idea of the mass renormalization.
A consequent QED formalism was developed by Dyson, Feynman, Schwinger, and Tomonaga.
They showed that all infinities can be removed from the
theory by so-called 
renormalization procedure. The basic idea of this procedure is the 
following. The electron mass and the electron charge, which originally
occur in the theory, are not directly measurable quantities. 
All physical quantities calculated
within QED become finite if they are expressed in terms of the
physical electron mass and charge,
parameters which can directly be measured in experiment.
All calculations in QED are based on the perturbation theory in the
fine structure constant $\alpha \approx 1/137.036$.
The individual
terms of the perturbation series are conveniently represented  by
so-called Feynman diagrams.

Before the beginning of 1970's investigations of QED effects in atomic
systems were mainly restricted to low-Z atoms such as hydrogen or helium
(here and below, $Z$ is the nuclear charge number).
In these systems, in addition to the small parameter $\alpha$, there
is another small parameter, which is $\alpha Z$. 
For this reason, all calculations of low-Z atoms were based
on the expansion in $\alpha$ and $\alpha Z$. 

A great progress
in experimental investigations of heavy few-electron ions, which was made for the
last decades (see \cite{bey97,bey03} and references therein),
has required accurate QED calculations for these systems.
Investigations of heavy few-electron ions play a special role in tests
of quantum electrodynamics. This is due to two reasons. First, in contrast
to low-Z atoms, the parameter $\alpha Z$ is not small and, therefore,
the calculations must be performed without any expansion in $\alpha Z$
\cite{moh98}.
Second, in contrast to heavy neutral atoms, the electron-correlation
effects can be calculated to a high accuracy using the pertubation theory
in the parameter $1/Z$. For this reason, the QED effects are not masked
by large electron-correlation effects, as it takes place in neutral atoms.
This provides an excellent opportunity to test QED 
 at strong electric fields.

The calculations of high-Z few-electron ions are generally based
on perturbation theory. To zeroth
order, one can consider that electrons interact only with the Coulomb
field of the nucleus. The interelectronic-interaction and QED effects
are accounted for by perturbation theory in the  parameters
$1/Z$ and $\alpha$,
respectively. This leads to quantum 
electrodynamics in the Furry picture. To formulate the perturbation
theory for calculations of the energy levels, transition and scattering
amplitudes, it is convenient to use the two-time Green function 
method \cite{sha02}.
For very heavy ions the parameter $1/Z$ becomes comparable
with $\alpha$ and, therefore, all the corrections may be classified 
by the parameter $\alpha$ only. In the present paper, we consider 
the current status of these calculations.

Relativistic units ($\hbar =c=1$) are used in the paper.

\section{Binding energies in heavy few-electron ions}

\subsection{ H-like ions}

To calculate the binding energy in
 a hydrogenlike ion, we may start with
the Dirac equation,
\begin{eqnarray} \label{dirac}
(\balpha \cdot {\bf p}+\beta m+V_{\rm C}(r))\psi({\bf r})=
E\psi({\bf r})\,,
\end{eqnarray}
where $V_{\rm C}(r)$ is the Coulomb potential induced
by the nucleus.
For the point-nucleus case, 
this equation leads to the binding energy
\begin{eqnarray} \label{bind}
E_{nj}-mc^2=-\frac{(\alpha Z)^2}{2\nu^2}\frac
{2}{1+(\alpha Z/\nu)^2 +\sqrt{1+(\alpha Z/\nu)^2}}\,mc^2\,,
\end{eqnarray}
where $\nu=n+\sqrt{(j+1/2)^2-(\alpha Z)^2}-(j+1/2)$,
 $n$ is the principal quantum number, and $j$ is 
the total angular momentum.
To get the binding energy to a higher accuracy,
we have to evaluate the quantum electrodynamic
and nuclear corrections.

The finite-nuclear-size correction can be obtained by 
solving the Dirac equation with the potential
induced by an extended-charge nucleus and taking the difference 
between the energies for the 
extended and the point nucleus model. 
This can be done either numerically (see, e.g.,
Ref. \cite{fra91}) or analytically \cite{sha93}.
To a relative accuracy of $\sim 0.2\%$,
in the range $Z$=1-100 the finite-nuclear-size correction 
is given by the following approximate formulas 
\cite{sha93}
\begin{eqnarray} \label{fns}
\Delta E_{ns}&=&\frac{(\alpha Z)^2}{10 n}[1+(\alpha Z)^2 f_{ns}(\alpha Z)]
\Bigl (2\frac{\alpha Z}{n}\frac{R}{(\hbar/mc)}\Bigr )^{2\gamma}mc^2\,,\\
\Delta E_{np_{1/2}}&=&\frac{(\alpha Z)^4}{40 }\frac{n^2-1}{n^3}
[1+(\alpha Z)^2 f_{np_{1/2}}(\alpha Z)]
\Bigl (2\frac{\alpha Z}{n}\frac{R}{(\hbar/mc)}\Bigr )^{2\gamma}mc^2\,,
\end{eqnarray}
where $\gamma=\sqrt{1-(\alpha Z)^2}$,
\be
f_{1s}(\alpha Z)&=&1.380-0.162\alpha Z+1.612(\alpha Z)^2\,,\nonumber\\
f_{2s}(\alpha Z)&=&1.508+0.215\alpha Z+1.332(\alpha Z)^2\,,\nonumber\\
f_{2p_{1/2}}(\alpha Z)&=&1.615+4.319\alpha Z-9.152(\alpha Z)^2
+11.87(\alpha Z)^3\,,\nonumber
\ee
and $R$ is an effective radius of the nuclear charge distribution
 defined by
\begin{eqnarray}
R=\Bigl\{\frac{5}{3}\langle r^2\rangle \Bigl [1-\frac{3}{4}
(\alpha Z)^2\Bigl (\frac{3}{25}\frac{\langle r^4\rangle}
{\langle r^2\rangle ^2}-\frac{1}{7}\Bigl)\Bigl]\Bigl\}^{1/2}\,.
\end{eqnarray}
The corresponding formulas for other states can be found 
in Ref. \cite{sha93}.
We note that, in contrast to the nonrelativistic
case where the corresponding correction 
is completely defined
by the root-mean-square radius, in heavy ions the 
higher-order moments of the nuclear charge distribution
are required  to determine  the nuclear-size correction 
on a 1\% accuracy level.  For instance, in case of Z=92
 to gain the precision $\sim$ 0.2\%
it is necessary to know the moment $\langle r^4 \rangle$.

The next corrections one should take into account
are the QED corrections of first order
in $\alpha$. They are determined
by the self-energy (SE) and vacuum-polarization (VP) diagrams 
(Fig. 1a,b). 
The contribution of the self-energy diagram (Fig. 1a)
 combined with the
corresponding mass counterterm is given by
\begin{eqnarray}
\Delta E&=&2i\alpha \int_{-\infty}^{\infty}d\omega
\int d\bfx_1\int d\bfx_2\; \psi_a^{\dag}(\bfx_1)
\alpha^{\mu} G(E_a-\omega,\bfx_1,\bfx_2)\nonumber\\ 
&&\times D_{\mu \nu}(\omega,\bfx_1-\bfx_2)
\alpha^{\nu}
\psi_a(\bfx_2)
-\delta m \int d\bfx \; \overline{\psi}_a(\bfx)\psi_a(\bfx)\,.
\end{eqnarray}
Here $\psi_a(\bfx)$ is the Dirac-Coulomb wave function of
the state under consideration, $\overline{\psi}=
\psi^{\dag}\gamma_0$,
 $G(\omega,\bfx_1,\bfx_2)$ is the Coulomb Green function,
$D_{\mu \nu}(\omega,\bfx_1-\bfx_2)$ is the photon propagator,
$\alpha^{\mu}=(1,\balpha)$, and $\balpha$ is a vector incorporating
the Dirac matrices.
The first evaluation of the SE correction for heavy ions
was performed by Desiderio and Johnson 
\cite{desid71} who employed the method suggested by Brown, Langer,
and Schaefer \cite{brown59}.
Later, Mohr \cite{moh74} developed another method which allowed him
to perform
a high precision evaluation of this correction in the range 
$Z=10-110$. An alternative approach to this problem
was worked out in Refs. \cite{sny91,blu91a,yer99a}.
 The most accurate calculations of the
 SE correction  to all orders in $\alpha Z$
were performed by Mohr \cite{moh92} and by Indelicato and Mohr
\cite{ind98} for the point nucleus case and by
 Mohr and Soff \cite{moh93} for the case of extended nuclei.
The highest accuracy for low-$Z$ ions was achieved
by Jentschura et al. \cite{jent99}.

The formal expression for the vacuum-polarization correction (Fig. 1b) is 
given by
\begin{eqnarray}
\Delta E=\frac{\alpha}{2\pi i}
 \int_{-\infty}^{\infty}d\omega
\int d\bfx_1\int d\bfx_2 \; \psi_a^{\dag}(\bfx_1)
\frac{1}{|\bfx_1-\bfx_2|}
[{\rm Tr}\, G(\omega,\bfx_2,\bfx_2)] 
\psi_a(\bfx_1)\,.
\end{eqnarray}
This expression is ultraviolet divergent. It can be
renormalized by dividing into two parts.
The first part corresponds
to the first nonzero term in the potential expansion of the
Coulomb Green function in powers of 
$\alpha Z$. This part, so-called Uehling part, becomes
finite due to the charge renormalization
and its evaluation causes no problem. 
The second part, so-called Wichmann-Kroll (WK) part,  
accounts for all higher-order terms
of the $\alpha Z$-expansion. Despite this part
 is finite, the regularization is still needed
 due to a spurious gauge-dependent piece of 
the light-by-light scattering contribution.
Calculations of the WK contribution in a wide range of
$Z$ were performed first by Soff and Mohr
\cite{soff88}
for the extended nucleus case and by Manakov et al.
\cite{manakov89} for the point nucleus case.
The most accurate results for some specific ions
were obtained by Persson et al. \cite{persson93}.

The QED corrections of second order in $\alpha$ 
are determined by diagrams presented in Fig. 2.
Most of these  diagrams can be calculated using 
the methods developed for the
first-order SE and VP corrections 
\cite{bei88,sch93,lin93,per96,mal96,bei97,plu98}.
The most demanding problem is to evaluate
the SE-SE diagrams  \cite{lab96} and the V(SE)P diagram (the diagram with 
a photon line inside the VP loop) \cite{zsch02}.
The loop-after-loop SE-SE diagram was first evaluated
by Mitrushenkov et al. \cite{mit95}.
 A partial evaluation of the other
 SE-SE diagrams was performed by Mallampalli and
Sapirstein \cite{mal98}. The residual SE-SE
terms were first calculated by Yerokhin and Shabaev \cite{yer01}.
Finally, in Refs. \cite{yer03}  the whole gauge invariant set 
of the SE-SE diagrams was evaluated in the range $Z=10-100$.
As to the S(VP)E diagram,  to date it was evaluated only in 
the Uehling approximation \cite{bei88,sch93}.

The calculations discussed above
are based on the approximation in which
 the nucleus is considered as a source
of the external Coulomb field. It defines
quantum electrodynamics within the external
field approximation.
The first step beyond this approximation would consist
in evaluation of the nuclear recoil correction.
As is known, in the nonrelativisic theory of a hydrogenlike
atom the recoil effect can easily be taken into account
 by using the electron reduced mass $\mu=mM/(m+M)$.
A rigorous relativistic theory of the recoil effect 
can be formulated only in the framework of QED.
In case of a hydrogenlike atom, a close formula 
for the recoil effect to first order in $m/M$
and to all orders in  $\alpha Z$
 was  derived in Ref. \cite{sha85}
(see also \cite{sha98a} and references therein). 
According to this formula,
the recoil correction  is given by the sum of a
low-order term $\Delta E_{\rm L}$ and a higher order
term $\Delta E_{\rm H}$, where
\begin{eqnarray}\label{rec1}
\Delta E_{\rm L}&=&\frac{1}{2M}\langle a|[
{\bf p}^2-
({\bf D}(0)\cdot{\bf p}+{\bf p}\cdot{\bf D}(0))
]|a\rangle\,,\\
\Delta E_{\rm H}&=&\frac{i}{2\pi M}\int_{-\infty}^{\infty}d\omega\,
\langle a|\Bigl({\bf D}(\omega)-
\frac{[{\bf p},V_{\rm C}]}{\omega+i0}
\Bigr)\nonumber\\
&&\times G(\omega +E_{a})
\Bigl({\bf D}(\omega)+\frac{[{\bf p},V_{\rm C}]}
{\omega+i0}\Bigr)
|a\rangle\,. \label{rec2}
\end{eqnarray}
Here ${\bf p}$ is the momentum operator,
 $G(\omega)$ is the Coulomb Green function,\\
$D_{m}(\omega)=-4\pi\alpha Z\alpha_{l}D_{lm}(\omega)\,,$
and $D_{ik}(\omega,r)$ is the transverse part of the photon
propagator in the Coulomb gauge.
In equation (\ref{rec2}),
the scalar product is implicit.
The term $\Delta E_{\rm L}$ contains all the recoil corrections
within the $(\alpha Z)^{4}m^2/M$ approximation. 
For the point nucleus case, it  can easily be calculated 
analytically \cite{sha85}
\begin{eqnarray}
\Delta E_{\rm L}=
\frac{m^{2}-E_{a}^{2}}{2M}\,.
\end{eqnarray}
The term $\Delta E_{\rm H}$ contains the contribution of order
$(\alpha Z)^{5} m^2/M$ and all contributions of higher order
in $\alpha Z$ which are not accounted for by the $\Delta E_{L}$
term.
Numerical evaluations of the recoil correction to all orders
in $\alpha Z$ were performed in Refs. 
\cite{art95,sha98b} for point and extended nuclei,
respectively.

Finally, one should take into account
the nuclear polarization correction,
which sets the ultimate limit of accuracy up to which QED 
can be tested in atomic systems. 
 This correction is determined by  the
electron-nucleus interaction diagrams in which
the intermediate states of the nucleus are excited. 
It was evaluated by Plunien and Soff 
\cite{plu95} and by Nefiodov et al. \cite{nef96}.

The individual contributions to the ground-state binding energy
 in $^{238}{\rm U}^{91+}$ are given in Table 1.
The uncertainty of the Dirac binding energy comes
from the uncertainty of the $R_{\infty}hc$
constant \cite{moh05}.
The finite-nuclear-size correction is evaluated for the Fermi
model of the nuclear charge distribution with 
$\langle r^2\rangle^{1/2}=5.8507(72)$ fm \cite{ang04}.
The uncertainty of this correction
 is estimated by adding quadratically two errors,
one obtained by varying the root-mean-square radius 
and the other obtained by changing the model of the nuclear-charge
distribution from the Fermi to the homogeneously-charged-sphere
model. 
As one can see from the table,
the present status of the theory and experiment 
on heavy H-like ions provides a test
of QED on the level of about 2\%.

\subsection{Li-like ions}

To date, the highest accuracy  
was achieved in experiments on the $2p_{1/2,3/2}-2s$ transitions
in heavy Li-like ions \cite{sch91,bei98,bos99,bra02}. 
In these systems, in addition to the one-electron 
contributions discussed above,
 one has to evaluate two- and three-electron contributions.
To first order in $\alpha$, the two-electron contribution
is determined by 
the one-photon exchange diagram (Fig. 3) whose
 calculation causes no problem.
To second order in $\alpha$, one should account for
the two-photon exchange diagrams (Fig. 4) and
the  self-energy and vacuum-polarization screening diagrams
 (Fig. 5). The two-photon
exchange diagrams were evaluated by different authors
\cite{yer00,sap01,and01,and03,art03}. 
The results of these calculations are
in a good agreement with each other. 
The complete calculation of the vacuum-polarization
screening diagrams  was performed in Ref. \cite{art99}.
The  self-energy screening
diagrams were evaluated in Refs. \cite{yer99,sap01}.
To gain an accuracy required by 
the experiments, in addition to higher-order one-electron
QED corrections,  one should  evaluate  
the interelectronic-interaction corrections of
third and higher orders in the parameter $1/Z$. 
Such evaluations within the framework of the Breit approximation
were accomplished in Refs. \cite{zhe00,sha01,and01}.

The individual contributions to
the $2p_{1/2}-2s$ transition energy in Li-like uranium are presented 
in Table 2.
The total theoretical value of the transition energy,
$280.66(22)$ eV, agrees well with 
the related experimental values, $280.59(10)$ eV
 \cite{sch91} and  $280.52(10)$ eV
 \cite{bra02}.
Comparing the first- and second-order QED contributions,
$-42.93$ eV and  $1.45(18)$ eV, respectively, 
with the total theoretical and experimental
uncertainties, we conclude that
the present status of the  theory for Li-like uranium
 provides a test of QED
on a $0.5$\% level to first order in $\alpha$ and 
on a $15$\% level to second order  in $\alpha$.

\subsection{He-like ions}

In Refs. \cite{mar95,gum04} the two-electron contribution 
to the ground state energy of heliumlike ions was measured.
It was done by comparing
the ionization energies of heliumlike and hydrogenlike ions.
This experiment is of special importance since to date the
two-electron contribution is the only measured value
which has been calculated to the second order in $\alpha$.
As in case of Li-like ions, the lowest-order two-electron
contribution is determined by the one-photon exchange diagram.
The contributions of second order in $\alpha$ are given
by the two-photon exchange diagrams and by the SE and VP screening
diagrams. The two-photon exchange contribution was first evaluated
by Blundell et al. \cite{blu93} and by Lindgren et al. 
\cite{lin95}. This contribution can be conventionally divided
into two parts: one which corresponds to the Breit approximation
and the other which is beyond the Breit approximation. 
The VP and SE screening diagrams for the ground state
 were evaluated in Refs.
\cite{per96a,yer97,art97,sun98}. The corresponding calculations
for excited states of He-like ions were performed
in Refs. \cite{art00,moh00b,lin01,and01,and03,and04,art05}.

In Table 3 we present the individual 
 contributions to the two-electron binding energy
of the ground state  in He-like uranium.
The two-photon exchange contribution 
 is divided into two parts as described above.
The Breit contribution  and the higher-order QED
corrections are evaluated as in Ref. \cite{art05}.
As one can see from the table, to test the screened
QED effects, 
  the experimental precision should be improved 
 by an order of magnitude.

\section{Hyperfine splitting in heavy ions}

High-precision measurements of the hyperfine 
splitting (HFS) in heavy hydrogenlike ions~\cite{exp1,exp2,exp3,exp4,exp5}
have triggered a great interest to
theoretical calculations of this effect.
The ground-state hyperfine splitting  of a hydrogenlike ion
is conveniently written as \cite{sha94hfs}
\begin{eqnarray}
\Delta E_{\mu}&=&\frac{4}{3}\alpha(\alpha Z)^{3}\frac{\mu}{\mu_{N}}
\frac{m}{m_{p}}\frac{2I+1}{2I}
mc^{2}\nonumber\\
& &\times\{A(\alpha Z)(1-\delta)(1-\varepsilon)+x_{\rm rad}\}\;.
\end{eqnarray}
Here $m_{p}$ is the proton mass,
$\mu$ is the nuclear magnetic moment, $\mu_{N}$ is the nuclear
magneton, and $I$ is the nuclear spin.
 $A(\alpha Z)$ denotes
the relativistic factor
\begin{eqnarray}
A(\alpha Z)=\frac{1}{\gamma(2\gamma-1)}=1+\frac{3}{2}(\alpha Z)^{2}
+\frac{17}{8}(\alpha Z)^{4}+\cdots \,,
\end{eqnarray}
% where $\gamma=\sqrt{1-(\alpha Z)^{2}}$.
 $\delta$ is the nuclear charge
 distribution
correction,
 $\varepsilon$ is the
nuclear  magnetization distribution correction (so-called 
Bohr-Weisskopf correction),
and $x_{\rm rad}$ is the QED correction.
The most accurate calculations of the QED corrections
were performed  in Refs. \cite{sha97,sun98a,art01}.
The uncertainty of the theoretical predictions is mainly determined
 by the uncertainty of the Bohr-Weisskopf (BW) effect. 
In calculations, based on 
the single-particle nuclear model~\cite{sha94hfs,lab95,sha97,sha99spr}, 
which provide a reasonable agreement
with the experiments (see, e.g., Ref. \cite{sha02}), 
this uncertainty may amount up to about 100\% of the BW effect
and is generally larger than the total QED contribution.
More elaborated calculations, based on many-particle nuclear
 models~\cite{tom95,sen02}, do not provide
a desirable agreement with the experiments. 

A new method to determine the BW effect was recently
developed in Ref. \cite{eli05}. 
In this method the BW correction  to the hyperfine
splitting in hydrogenlike $^{209}$Bi, $^{203}$Tl, and $^{205}$Tl 
was determined using experimental data on the hyperfine splitting in the 
corresponding muonic atoms \cite{expmu1,expmu2}.
The parameters of the nuclear magnetization distribution were chosen
to reproduce the experimental values of the nuclear magnetic moment
as well as the BW effect in  muonic atoms extracted from the corresponding
experiments. The single-particle and configuration-mixing nuclear models
were considered. To increase the precision of determining the BW contribution,  
the QED corrections for muonic atoms have been evaluated. 
In Table 4, the BW correction to the HFS in 
electronic H-like ions, derived from the 
experiments on muonic atoms, is compared with the BW correction
obtained by direct calculations within the single- and many-particle
nuclear models. Taking into account that the uncertainty of
the single-particle results may amount up to 30-100\%,
we conclude that the $\varepsilon$ values, based on experiments
with muonic atoms, have a better accuracy.

In Table \ref{ta:ok}, we compare  the total theoretical results for the HFS in 
 H-like ions  with experiment.
As one can see from the table, the results based on experiments
with muonic atoms
are closer to the experimental ones, 
compared to the results based on the direct calculations 
within the single-particle nuclear model.
However, due to a higher accuracy of the former results, a small 
discrepancy between the theory and experiment occurs for 
$^{209}$Bi$^{82+}$. It can also be seen that in case of two isotopes of Tl the
results of Ref. \cite{eli05} are much closer to the related experimental
values than in case of Bi. Since, in contrast to the Tl isotopes,
the bismuth nucleus has a nonzero electric-quadrupole moment, one may expect
that the discrepancy between the theory and experiment is caused by a large
contribution of the electric-quadrupole splitting in 
muonic bismuth. It is known that
the first-order electric-quadrupole splitting 
 vanishes for $s$ states. However, the second-order effect
 is nonzero and, in principle, it may be significant
for muonic atoms. This is due to a relatively large role
of the electric-quadrupole HFS interaction in muonic atoms compared to
electronic ones.
Our numerical evaluation of this effect showed that, whereas
it changes the HFS in muonic bismuth by about $-$3.4 eV,
it almost does not affect the BW value presented in Table 4. 

One of the main goals of the HFS experiments with heavy H-like ions
was to probe the magnetic sector of QED in the presence of a strong
Coulomb field. The analysis of the theoretical results showed, however,
that the QED corrections to the HFS in heavy H-like ions are
strongly masked by the uncertainty of the BW effect. This makes
unfeasible tests of QED by a direct comparison of theory and experiment 
on the HFS in heavy H-like ions. An opportunity for QED tests
has been found by considering a specific difference of the
ground-state HFS values in H- and Li-like ions \cite{sha01hfs}. 
Namely, it was shown that the difference
\begin{eqnarray} \label{delprime}
\Delta'E=\Delta E^{(2s)}-\xi \Delta E^{(1s)}\,,
\end{eqnarray}
where $\Delta E^{(1s)}$ and $\Delta E^{(2s)}$ are
 the HFS in H- and Li-like ions of the same isotope,
is very stable with respect to variations of the nuclear 
model, if the parameter $\xi$ is chosen to cancel
the BW corrections in the right-hand side of equation (\ref{delprime}). 
The parameter $\xi$ is almost independent of the nuclear
structure and, therefore, can be calculated to a high accuracy.
In case  of Bi, the calculations yield $\xi$ = 0.16885
and  $\Delta'E$= $-$61.27(4).
The non-QED and QED contributions amount to
-61.52(4) and 0.24(1), respectively.
It means that the QED contribution 
is six times larger than the current total theoretical uncertainty.  
 This provides good perspectives for tests of QED
in the HFS experiments. In particular, it can be shown that 
this method allows one to test  QED on 
level of a few percent, provided 
the HFS is measured to accuracy $\sim 10^{-6}$.

\section{Bound-electron g-factor}

The g-factor of a hydrogenlike ion with a spinless nucleus
 can be defined as
\be
g^{(e)}=-\frac{\langle JM_{J}|\mu_{z}^{(e)}|J M_{J}\rangle}
{\mu_{B} M_{J}}\,,
\ee
where $\bmu^{(e)}$ is the operator of the magnetic
moment of electron and $\mu_{B}$ is the Bohr magneton.
For the $1s$ state,
a simple relativistic calculation based on the Dirac 
equation yields
\be
g_{\rm D}=2-(4/3)(1-\sqrt{1-(\alpha Z)^2})\,.
\ee
The total g-factor value can be written as
\be
g^{(e)}=g_{\rm D}+\Delta g_{\rm QED}+\Delta g_{\rm rec}
+\Delta g_{\rm NS}\,,
\ee
where $\Delta g_{\rm QED}$ is the QED correction, 
$\Delta g_{\rm rec}$ is the nuclear recoil correction,
and $\Delta g_{\rm NS}$ is the finite-nuclear-size
correction. High-precision measurements  of the bound-electron
 g-factor in H-like carbon \cite{her00,haf00} and oxygen
\cite{ver04} have stimulated successful theoretical
calculations of this effect 
(see Refs. \cite{bei00,yer02,sha02a,sha02b,mar01,nef02,pac05}
and references therein). In particular, these studies 
provided a new determination of the electron mass
to an accuracy which is four times better than that of
the 1998
CODATA value. As a result, the 2002 CODATA value for the electron
mass \cite{moh05} is derived from the 
g factor measurements. An extension of these experiments
to higher-Z systems, which is anticipated in the near
future, could lead also to an independent determination of 
the fine structure constant.

For heavy H-like ions the theoretical uncertainty of
the bound-electron g-factor is mainly determined 
by the nuclear effects \cite{sha02b,nef02}.
This uncertainty becomes comparable with the
binding QED corrections of second order in $\alpha$
and, therefore, strongly restricts probing QED
in these investigations. However, in Ref. \cite{sha02c}
it was shown that the uncertainty caused by the nuclear effects
can be significantly reduced in a specific difference
 of the g-factors of H- and Li-like ions of the same 
isotope. This gives a good opportunity for tests of the
magnetic sector of QED in the presence of a strong Coulomb
field. The most accurate calculations of the g-factor
of Li-like ions in the range Z=6$-$92 were performed
in Ref. \cite{gla05}.

The g-factor of a H-like ion with  nonzero nuclear spin
can be approximated as 
\begin{eqnarray}\label{gat}
g&=& g^{(e)}\frac{F(F+1)+j(j+1)-I(I+1)}{2F(F+1)}\nonumber\\
&&-\frac{m}{m_p}g_I\frac{F(F+1)+I(I+1)- j(j+1)}{2F(F+1)}\,.
\end{eqnarray}
Here $m$ and $m_p$ are the electron and the proton mass,
respectively, $g^{(e)} $ is the  bound-electron g-factor
defined above,  $g_I=\mu/\mu_n I$ is the nuclear
g-factor, $\mu_n={|e|}/{2m_p c}$ is the nuclear magneton,
$j$ and $ I$ are the electron and the nucleus angular momentum,
respectively, and $F$ is the total angular momentum of the ion.
According to formula (\ref{gat})
 the contribution of the nuclear g-factor is suppressed
by factor $m/m_p$ compared to the electronic g-factor.
It follows that measurements
 of the g-factor of ions with nonzero nuclear spin with a
 $10^{-9}$ accuracy 
 would provide determinations
of the nuclear magnetic moments on the $10^{-6}$ accuracy level.
 Calculations of various
corrections to formula  (\ref{gat})
 were presented in Ref. \cite{mos05}.

Another possibility for investigations of the
g-factor of ions with nonzero nuclear spin
 was  discussed in  Ref. \cite{sha98can}.
In that work it was shown that the transition
probability between the ground state hyperfine splitting
components of a hydrogenlike ion, including the
QED and nuclear corrections, is given by
\begin{eqnarray} \label{trans}
w=\frac{\alpha}{3}\,\frac{\omega^3}{m^2}\,\frac{I}{2I+1}
\Bigl[g^{(e)}-g_{I}\frac{m}{m_{p}}\Bigr]^2\,,
\end{eqnarray}
where $\omega$ is the transition frequency.
Formula (\ref{trans}) allows one to
 calculate  the QED and nuclear
corrections to the transition probability using
 the corresponding
corrections to the bound-electron g-factor.
In particular,
 it was found \cite{sha98can}
that in  
cases of Pb and Bi the
 QED and nuclear corrections increase the transition probability by
about 0.3\%. In Ref. \cite{win98} the life time of the upper
hyperfine splitting component in $^{209}{\rm Bi}^{82+}$
was measured to be $\tau_{\rm exp}$=397.5(1.5) 
$\mu$s. This result is in a good agreement 
with the theoretical prediction \cite{sha98can}
$\tau_{\rm theo}=399.01(19) \mu$s.
Using formula (\ref{trans}) and the experimental values of the
hyperfine splitting and the transition probability in
$^{209}{\rm Bi}^{82+}$ \cite{exp1,win98},
one finds the experimental value of the bound-electron g-factor
in Bi to be 1.7343(33). The corresponding
theoretical value  is  1.7310.
 The values of the individual contributions
 to the bound-electron g-factor in Bi are given in Table 6.
From this table, it can be seen
 that including the QED correction
is needed to obtain the agreement between 
the theory and experiment.

\section{Parity nonconservation effects with heavy ions}

Investigations of parity nonconservation (PNC) effects in atomic systems
play a prominent role in tests of the Standard Model (SM) 
\cite{khr91}.
 The well-known cesium experiment by
Wieman's group \cite{woo97}, compared to the most elaborated theoretical 
result
(see Ref. \cite{sha05} and references therein), 
provided
the most accurate test of electroweak theory at the low-energy regime.
Further improvement of tests of the Standard Model with neutral atoms,
from theoretical side, is mainly limited by difficulties with accurate
calculations of the electron-correlation effects. For this reason, the PNC
experiments with heavy few-electron ions, where the electron-correlation
effects can be evaluated to a high accuracy, seem highly desirable.
The PNC effects in these systems were first discussed in Refs.
\cite{gor74,sch89,kar92}. For tests of the spin-independent part
of the week interaction, a promising situation occurs
in heavy He-like ions with the nuclear charge number  near $Z=64$
and $Z=92$, where two levels of the opposite parity,
$2^1S_0$ and $2^3P_0$, are very close to each other.
The study of PNC effects with these ions
 requires precise knowledge of the $2^1S_0 - 2^3P_0$ energy
difference. The most accurate calculations of this difference,
which include a complete set of two-electron 
QED corrections,
were performed in Ref. \cite{art05}. In Table 7, the results of
Ref. \cite{art05} are compared with the related
data by other authors.

A feasible PNC experiment 
with heavy ions was suggested by Labzowsky et al. \cite{lab01}.
Instead of the standard measurement of the circular dichroism which
is rather difficult to perform with x-ray radiation, it was proposed 
to study a quenching-type experiment with interference of hyperfine-
and weak-quenched transitions in polarized He-like europium.
The idea of the experiment is the following. The basic
one-photon decay  $2^1S_0 - 1^1S_0$
channel is  the hyperfine-quenching (hfq) M1 transition 
 which is due to the
hyperfine mixing of the $2^1S_0$ and  $2^3S_1$ levels.
Another one-photon decay  $2^1S_0 - 1^1S_0$  channel 
is due to the mixing of  the $2^1S_0$ and  $2^3P_0$ levels
caused by the weak interaction of electrons with the nucleus
and due to the hyperfine mixing of  the $2^3P_0$ and  $2^3P_1$ levels.
As a result, the total amplitude of the one-photon $2^1S_0 - 1^1S_0$
transition is a mixture of the basic M1 and the additional E1 amplitude.
The polarization of  Eu$^{61+}$ in the  $2^1S_0$ state can be
described by the density matrix \cite{lab01}
\begin{eqnarray}
\rho = \frac{1}{2I+1}\Bigl[1+\frac{3\lambda_0}{I+1}(\bnu\cdot {\bf I})\Bigr]\,,
\end{eqnarray}
where ${\bf I}$ is the operator of the total angular momentum,
$\bnu$ is the unit vector directed along the ion polarization,
and $\lambda_0$ is the degree of polarization $(\lambda_0\le 1)$.
The probability for the emission of a photon
in direction ${\bf n}$ is given by
 \cite{lab01}
\begin{eqnarray} \label{dW}
dW({\bf n})=\frac{W_{\rm M1}}{4\pi} [1+\varepsilon(\bnu\cdot {\bf n})]d\Omega\,,
\end{eqnarray}
where  $W_{\rm M1}$ is the total probability of
the hfq M1  $2^1S_0 - 1^1S_0$ transition and
$\varepsilon$ is the asymmetry coefficient caused by the PNC effects.
Our evaluation of this coefficient yields
\begin{eqnarray} \label{asym}
\varepsilon = 6\lambda_0\sqrt{W_{\rm E1}/W_{\rm M1}}\,/(I+1)\,,
\end{eqnarray}
where $W_{\rm E1}$ is the total probability of
the weak-hyperfine quenching E1 $2^1S_0 - 1^1S_0$ transition.
We note that expression (\ref{asym}) differs by about 
a factor of 2 from the
related expression derived in Ref. \cite{lab01}.

The experiment should consist in observing the difference 
of the transition probability (\ref{dW}) due to a change
of the ion polarization direction or, equivalently, due to
rotating the detector around the beam direction by an angle $\pi$.
The value of this difference is proportional to the asymmetry parameter
$\varepsilon$. Our calculation of this parameter for europium
employing the transition energies from Ref. \cite{art05}
(see Table 7)
yields $\varepsilon \approx 0.0004\lambda_0$. This value is almost
four times larger compared to that obtained in
Ref. \cite{lab01}. This
is due to a change 
of the $2^1S_0 - 2^3P_0$  energy
difference, compared to that used in Ref. \cite{lab01}, 
and due to the change of the expression for $\varepsilon$
discussed above.

The experiment under consideration requires preparing
and storing a polarized ion beam. A promising idea
for preparation of polarized ion beams 
was suggested in Ref. \cite{pro03}.

\section*{\large {\bf Conclusion}}

In this paper we have considered the present status of
calculations of the QED effects in heavy  
few-electron ions. Calculations of the PNC effects
for heavy ions have also been discussed.

To date, the most accurate tests of QED effects on binding
energies in a strong Coulomb field have been accomplished 
in heavy Li-like ions:  on a 0.5\% accuracy level to first 
order in $\alpha$ and on a 15\% accuracy level to second
order in $\alpha$. An improvement of the experimental
accuracy by an order of magnitude accompanied by 
more accurate determinations of higher-order QED 
and nuclear-size effects
would provide tests of QED  
beyond the external field approximation.

High-precision measurements of the hyperfine splitting 
in heavy H-and Li-like ions of the same isotope are highly 
desirable. They would give
a unique opportunity for tests of the magnetic sector of QED
in the strongest electric field currently available
for experimental study.

The QED theory of the bound-electron g-factor has
been probed by direct measurements on hydrogenlike
carbon and oxygen. These measurements have also provided
a new determination of the electron mass with an accuracy
which is four times better than that of the previously accepted value.
Extentions of these measurements to higher-$Z$ systems and
to ions with nonzero nuclear spin would
 provide independent determinations of the
fine structure constant and the nuclear magnetic moments.

Investigations of the PNC effects with heavy ions seem very
promising for tests of the Standard Model.

\section*{\large {\bf Acknowledgements}}

This work was supported in part by RFBR (Grant No. 04-02-17574) and
by INTAS-GSI (Grant No. 03-54-3604). A.N.A., S.S.B., Y.S.K.,
N.S.O., and V.A.Y
acknowledge  the support by the ``Dynasty'' foundation.

\newpage

\begin{table}
%\tablewidth=8.5cm
\centering
\caption{Individual contributions to the
ground-state
binding energy 
in $^{238}{\rm U}^{91+}$, in eV.
 The Lamb shift is defined as a part of the binding
energy that is beyond its point-nucleus value
 given by equation (\ref{bind}).}
\vspace{0.2cm}
\tabcolsep=.6cm
\begin{tabular}{l c } \hline
Point-nucleus binding energy & -132279.93(1)  \\ 
Finite nuclear size  
 &       198.33(52) \\ 
First order SE   & 355.05 \\
First order VP   & -88.60 \\
Second-order QED   &  -1.26(33)\\ 
Nuclear recoil  &  0.46   \\ 
Nuclear polarization      &    -0.20(10)\\
Lamb shift theory  &  463.78(62)   \\
Lamb shift experiment \cite{gum05}   &460.2(4.6)\\
 \hline
\end{tabular}
\end{table}

\begin{table}
%\tablewidth=8.5cm
\centering
\caption{The $2p_{1/2}-2s$ transition energy  
 in $^{238}{\rm U}^{89+}$, in eV.}
\vspace{0.2cm}
\tabcolsep=.6cm
\begin{tabular}{l c } \hline
One-photon exchange &     368.83 \\
One-electron nuclear size   & -33.27(8)  \\ 
First-order QED   & -42.93\\
Two-photon exchange within & \\
the Breit approximation & -13.54\\
Two-photon exchange beyond & \\
the Breit approximation &   0.17\\ 
SE and VP screening & 1.16 \\
Three- and more photon exchange   & 0.16(7)\\
Nuclear recoil  &-0.07\\
Nuclear polarization    &  
  0.03(1)\\
One-electron second-order QED & 0.12(18)\\
Total theory  &  280.66(21)    \\
Experiment \cite{sch91}   &280.59(10)\\
Experiment \cite{bra02}   &280.52(10)\\
 \hline
\end{tabular}
\end{table}

\begin{table}
%\tablewidth=8.5cm
\centering
\caption{The two-electron binding energy of the ground state 
 in  $^{238}{\rm U}^{90+}$, in eV.}
\vspace{0.2cm}
\begin{tabular}{l c } \hline
One-photon exchange & 2265.90(1)  \\ 
Two- and more photon exchange & \\
within the Breit approximation  &     $-$11.96\\ 
Two-photon exchange & \\
beyond the Breit approximation & $-$0.85   \\
SE screening & $-$9.78 \\
VP screening & 2.63 \\
Higher-order QED & $-$0.05(18) \\
Total theory   & 2245.89(18)   \\
Experiment \cite{gum04} & 2248(9)\\
 \hline
\end{tabular}
\end{table}

\begin{table}
	\caption{The Bohr-Weisskopf correction  $\varepsilon$,
 derived from experiments on muonic atoms \cite{eli05}, 
is compared with previous direct evaluations of this effect
within the single-particle \cite{lab95,sha99spr} and the
many-particle \cite{tom95,sen02} nuclear model.}\label{ta:bw}
\vspace{0.1cm}
	\begin{center}
	\begin{tabular}{llll}
	\hline
	               &$^{209}$Bi$^{82+}$&$^{205}$Tl$^{80+}$&$^{203}$Tl$^{80+}$\\
	\hline
Elizarov et al. \cite{eli05}       &0.0123(15)       
 &0.0193(27)        &0.0155(40)\\
	Shabaev ~\cite{sha99spr} &0.0118    
        &0.0179            &0.0179     \\
	Labzowsky et al. ~\cite{lab95}  &0.0131     
       &                  &           \\
	Tomaselli et al.  ~\cite{tom95} &0.0210     
       &                  &           \\
	Sen'kov and Dmitriev ~\cite{sen02}  
&0.0095($+$7, $-$38)&&                \\           
	\hline
	\end{tabular}
	\end{center}
\end{table}

\begin{table}
\caption{Hyperfine splitting in  H-like ions, in eV.}\label{ta:ok}
\vspace{0.1cm}
\begin{center}
\begin{tabular}{cccc}
	\hline
	                     &  Theory \cite{eli05}    &Theory \cite{sha99spr}
    & Experiment \\
	\hline
	$^{203}$Tl$^{80+}$   & 3.220(20)   & 3.229(17) 
& 3.21351(25) \cite{exp5}                     \\
	$^{205}$Tl$^{80+}$   & 3.238(9)    & 3.261(18) & 3.24410(29) \cite{exp5}\\
	$^{209}$Bi$^{82+}$   & 5.098(7)    & 5.101(27) & 5.0840(8) \cite{exp1}\\
	\hline
\end{tabular}
\end{center}
\end{table}

 \begin{table}
\caption{The bound-electron g-factor in  $^{209}{\rm Bi}^{82+}$.}
\vspace{0.1cm}
\begin{center}
\begin{tabular}{ll} \hline
Point-nucleus Dirac value& 1.7276 \\
QED&0.0029\\
Nuclear size correction &0.0005\\
Total theory & 1.7310\\
Experiment \cite{win98} & 1.7343(33)\\ \hline
\end{tabular}
\end{center}
\end{table}

 \begin{table}
\caption{The $2\,^3P_0-2\,^1S_0$ transition energy, in eV.}
\vspace{0.2cm}
\begin{center}
\begin{tabular}{ccccc} \hline
$Z$ & Artemyev et al. & Andreev et al. &
Plante et al. & Drake \\
   & \cite{art05} & \cite{and03}&  \cite{pla94} & \cite{dra88}\\
\hline
63 & -0.224(69) & -0.591 &       & -0.168 \\
64 & 0.004(74) & -0.389 & -0.170 & 0.067 \\
65 & 0.32(12) &-0.153 &         & 0.328 \\
66 & 0.495(84) & 0.016 & 0.341 & 0.614 \\
89 & 1.61(46) & & & 1.731 \\
90 & 0.61(46) & & -0.095& 0.718 \\
91 & -0.31(55) & -1.971 & & -0.209 \\
92 & -2.64(28) &-4.511 & -2.639 & -1.816 \\
\hline
\end{tabular}
\end{center}
\end{table}

\newpage

\begin{figure} \caption{First-order self-energy and vacuum-polarization
diagrams.}
\vspace{0.2cm}
\centerline{ \mbox{
\epsfysize=3cm \epsffile{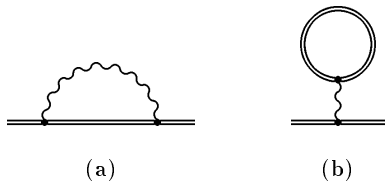}
}}
\end{figure}

\begin{figure} \caption{Second-order one-electron Feynman diagrams.}
\vspace{0.2cm}
\centerline{ \mbox{
\epsfysize=8cm \epsffile{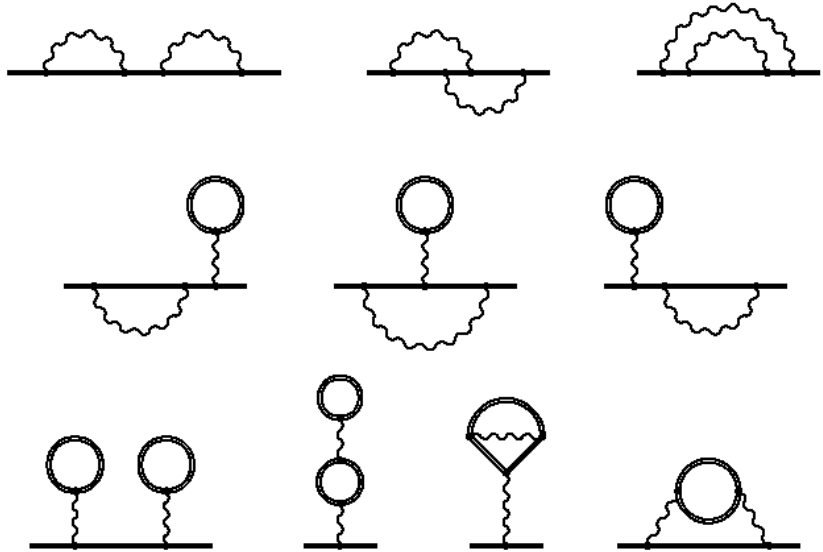}
}}
\end{figure}

\begin{figure} \caption{One-photon exchange diagram.}
\vspace{0.2cm}
\centerline{ \mbox{
\epsfysize=2,5cm \epsffile{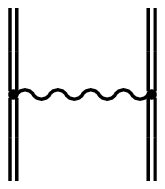}
}}
\end{figure}

\begin{figure} \caption{Two-photon exchange diagrams.}
\vspace{0.2cm}
\centerline{ \mbox{
\epsfysize=3.5cm \epsffile{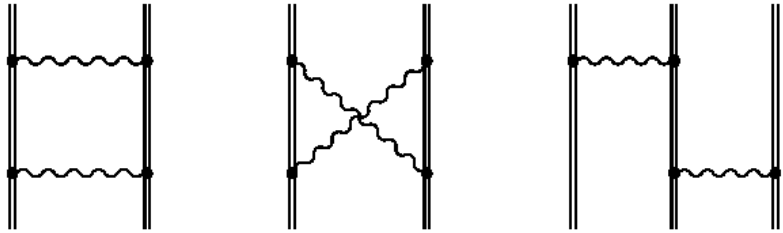}
}}
\end{figure}

\begin{figure} \caption{Self-energy and vacuum-polarization
 screening diagrams.}
\vspace{0.2cm}
\centerline{ \mbox{
\epsfysize=5cm \epsffile{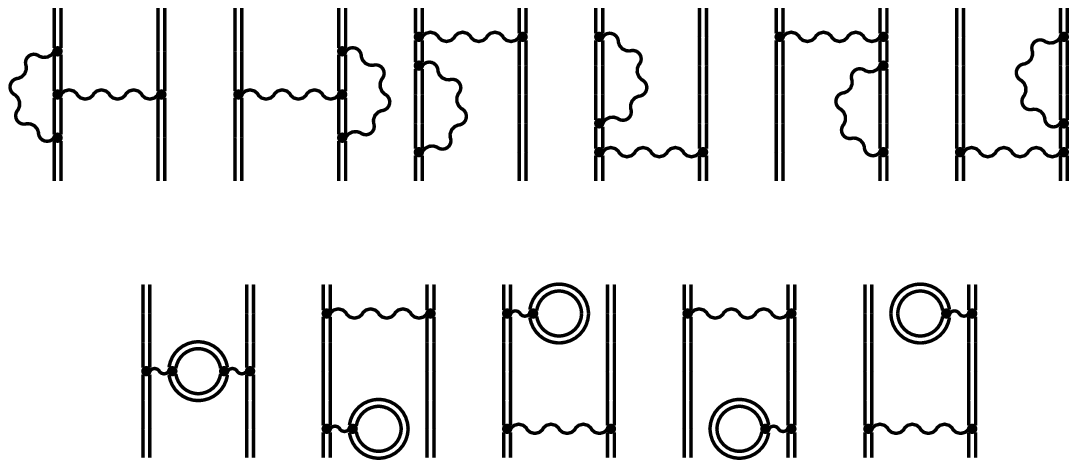}
}}
\end{figure}

\end{document}